\documentclass[aps,prb,twocolumn,floatfix,floats,showpacs,superscriptaddress]{revtex4}
\usepackage{graphicx,latexsym}
\usepackage{dcolumn}
\usepackage{amsmath}

\begin{document}
\title{Coherent quantum transport in narrow constrictions in the presence of \\
a finite-range longitudinally polarized time-dependent field}

\author{C. S. Tang}
\affiliation{Physics Division, National Center for Theoretical
        Sciences, P.O.\ Box 2-131, Hsinchu 30013, Taiwan}
\author{C. S. Chu}
\affiliation{Department of Electrophysics, National Chiao Tung
University, Hsinchu 30010, Taiwan}
\begin{abstract}
We have studied the quantum transport in a narrow constriction acted
upon by a finite-range longitudinally polarized time-dependent
electric field. The electric field induces coherent inelastic
scatterings which involve both intra-subband and inter-sideband
transitions. Subsequently, the dc conductance $G$ is found to exhibit
suppressed features. These features are recognized as the
quasi-bound-state (QBS) features which are associated with electrons
making transitions to the vicinity of a subband bottom, of which the
density of states is singular. Having valley-like instead of dip-like
structures, these QBS features
are different from the $G$ characteristics for constrictions  acted upon
by a  finite-range time-modulated potential.  In addition, the subband
bottoms in the time-dependent electric field region are
shifted upward by an energy proportional to the square of the electric
field and inversely proportional to the  square of the frequency.
This effective potential barrier is originated from the square of the
vector potential and it leads to the interesting field-sensitive QBS
features.  An experimental set-up is proposed for the observation of
these features.
\end{abstract}

\maketitle

Quantum transport in mesoscopic systems has received much
attention in recent years.
Among the most studied mesoscopic structures is the quantum point
contact (QPC), due to its simple configuration.
In such QPC systems, the lateral energy is quantized into subbands,
giving rise to a quantized conductance $G$.~\cite{wee88,wha88,wee88b}
These QPCs can be created electrostatically by negatively
biasing the  split-gates located on top of a ${\rm
GaAs-Al_{x}Ga_{1-x}As}$ heterostructure.~\cite{wee88,wha88,wee88b}

\begin{figure}[b]
      \includegraphics[width=0.44\textwidth,angle=0]{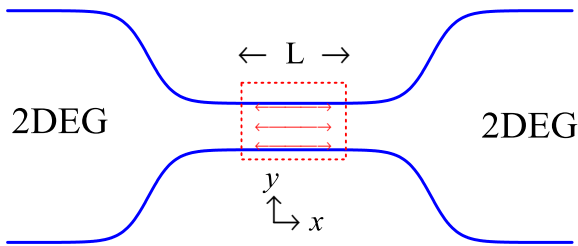}
      \caption{Sketch of the gated QPC which is connected at each end to a
two-dimensional electron gas electrode. The narrow constriction is
acted upon  by an external longitudinally polarized time-dependent
electric field within millimeter wave range.} \label{fig:1}
\end{figure}

The QPC can be pictured as a narrow constriction
connecting adiabatically at each end to a two-dimensional electron
gas (2DEG),~\cite{gla88,gla90} as depicted in Fig.~\ref{fig:1}.    The
energy levels in the narrow constriction are  quantized into
one-dimensional subbands which density of states (DOS) is singular at
a subband bottom. In the presence of an attractive scatterer, such
singular DOS gives rise to dip structures in $G$,
\cite{chu89,bag90,tek91,nix91,lev92,tak92,kun92}
which is associated with the impurity-induced QBSs~\cite{bag90} formed
just beneath a subband bottom.

Recently, there are growing interest in the time-dependent
responses of  QPC structures. The time-dependent fields that act
upon the QPCs can be transversely
polarized,~\cite{hek91,hu93,wys93,fed93,jan94,gor94,gri95,maa96,%
chu96,hu96,tag96,tag97} longitudinally polarized,~\cite{fen93} or
without polarization but represented by time-dependent
potentials.~\cite{bag92,tan96} In all these studies, the region
acted upon by the time-dependent fields has dimensions  shorter than
the incoherent mean free path,  so that electrons undergo
predominately coherent inelastic scatterings within the region.
Several interesting effects have been studied. First, the mechanism
of time-modulated electron pumping has been proposed\cite{hek91} in
an unbiased asymmetric QPC which is acted upon by a time-dependent
transverse field. Second, the effects of photon-assisted processes
on the quantum transport is
studied.~\cite{hu93,wys93,fed93,jan94,gor94,gri95,maa96} The QPC
considered is unbiased, has varying widths, and is acted upon by a
time-dependent transverse field. More recently,  QBS features are
predicted to occur in a narrow constriction which is acted upon by a
time-dependent potential.~\cite{bag92,tan96} These QBS features are
of similar physical origin as the the impurity-induced
QBS.~\cite{chu89,bag90,tek91,nix91,lev92,tak92,kun92}

Several methods have been developed to explore the time-modulated
phenomena in QPCs. Earlier attempts expanded the scattering
wavefunction in terms of the adiabatic wavefunctions of the
QPC,~\cite{hek91,fen93} and converted the time-dependent Schr\"{o}dinger
equation into a set of coupled differential equations for the
expansion coefficients.
In striving for an analytic result, the authors had to
ignore processes other than one- or two-photon processes. Later on,
a hybrid recursive Green's function method has been proposed.
\cite{gri95,maa96}  The formalism allows, in principle,
multi-photon processes to be incorporated systematically and
numerically up to many orders.  However, their calculation
has neglected the $A^2$ term in the Schr\"{o}dinger equation,
where ${\bf A}$ is the vector potential for the time-dependent
electric field. Thus intrinsic two-photon processes
have been neglected.
Recently, a transfer-matrix method has
been devised to study the effect of a longitudinally polarized
time-dependent electric field on photon-assisted tunneling in double
barrier structures.~\cite{wag97}
The potential that represents the electric field was sliced into
piece-wise uniform potentials, and the method involved cascading
the transfer matrices due to each of these piece-wise uniform
potentials. The focus of the work is on the scaling characteristics
of the photon frequency in the photon-assisted tunneling.
In this work, we focus on the QBS characteristics instead.  By
utilizing a vector potential to represent the electric field, we
propose a matching scheme that avoids slicing the region covered by
the electric field. Thus our method is efficient and allows detail
analysis of the QBS features.

We note here that the situations we considered are different from
those in which the source and the drain electrodes are biased
with an ac voltage.~\cite{but93,chr96,aro97}
In the latter situations, the time-dependent fields cover a region
including also the reservoirs, which then is of dimension
longer than the incoherent mean free path. Whether the
QBS features should remain robust in these latter situations is an
interesting issue, and is left to further study.

In this paper, we study the quantum transport
in a narrow constriction which is acted upon by an
external time-dependent electric field.  This
electric field is chosen to be polarized longitudinally,
with ${\bf E}({\bf x},t)={\cal E}_{0}\cos (\omega t) \Theta (L/2 -
|x|)\hat{x}$, where $L$ denotes  the  range covered by the field and
$x$ represents the transmission direction.  The finiteness
in the range of the electric field
breaks the translational invariance, hence allows the
coherent inelastic scatterings  not to conserve the longitudinal
momentum.
\cite{chu96,tan96} On the other hand, the uniformity of the electric
field in the transverse direction allows only intra-subband
transitions in the inelastic scattering processes, and leaves
the subband index $n$ intact.

In the following we select the energy unit $E^{*}=\hbar^{2}
k_{F}^2/(2m^*)$, the length unit $a^*=1 / \! k_F$, the time
unit $t^*=\hbar / E^*$, the frequency unit $\omega^* = 1/t^*$, and
the field amplitude ${\cal E}_0$ in units
of $E^{*}/(ea^{*})$. Here $-e$ denotes the charge of an electron,
$m^*$ represents the effective mass, and $k_F$ is a typical Fermi wave
vector of the reservoir.  The time-dependent Hamiltonian is of
the form ${\cal H} = {\cal H}_{y} + {\cal H}_{x}(t)$, where
${\cal H}_{y} = -\partial^{2}/\partial y^{2} + \omega_{y}^{2} y^{2}$
contains a quadratic confinement. The quantized transverse
energy levels are $\varepsilon_{n} = (2n+1)\omega_{y}$, and
the corresponding wavefunctions are denoted by $\phi_{n}(y)$.
\cite{but90} The time-dependent part of the Hamiltonian ${\cal
H}_x (t)$  is given by
\begin{equation}
{\cal H}_x (t) = - \left[
{\partial \over \partial x}-{i{\cal E}_0\over \omega} \sin (\omega t)
\Theta \left( {L\over2} - |x| \right)
\right]^2\, ,
\label{eq:hxt}
\end{equation}
where the electric field is represented by a vector potential.
  For an $n$th subband electron incident upon the time-modulated
region from the left reservoir with a total energy $\mu$,
the scattering wavefunctions are of the form
$\Psi^{+}_n({\bf x},t)=\psi(x,t,\mu_n)\phi_n(y)
\exp(-i\mu_n t)$, where $\mu_{n} = \mu - \varepsilon_{n}$, and the
subband index remains
unchanged. The longitudinal part $\psi(x,t,\mu_n)$ of the scattering
wavefunction can be casted into the form
\begin{eqnarray}
\psi(x,t,\mu_{n}) &=& \exp\left[
{i{\cal E}_0^2\over 4\omega^3}\sin(2\omega t)\right]
\nonumber \\
&\times& \exp\left[ {-2{\cal E}_0\over \omega^2}
\cos(\omega t){\partial \over\partial x}
\right] \varphi(x,t,\mu_n)\, ,
\end{eqnarray}
which, when being substituted into Eq.~(\ref{eq:hxt}) in the
time-modulated region, would lead to an effective Schr\"odinger
equation for $\varphi(x,t,\mu_n)$, given by
\begin{equation}
i{\partial\over \partial t}\varphi(x,t,\mu_n) =
\left[ -{\partial^2\over \partial x^2}
+ {{\cal E}_0^2\over 2\omega^2}
\right] \varphi(x,t,\mu_n)\, .
\label{eq:ippt}
\end{equation}
The effective Schr\"{o}dinger equation Eq.~(\ref{eq:ippt}) satisfied by
$\varphi(x,t,\mu_n)$ has a constant potential term ${\cal E}_0^2 /
2\omega^2$ which is from the square of the vector potential.  This
suggests that the electric field causes an effective potential barrier
in the  time-modulated region, and this potential barrier depends on
the amplitude and the frequency of the electric field.
Making use of the complete set of the $\varphi(x,t,\mu_n)$ functions,
we can write down the general form of the longitudinal wavefunction
$\psi(x,t,\mu_n)$ in the time-modulated region.  Along the entire
$x$-axis, the wavefunction $\psi(x,t,\mu_n)$ can be written in the
following form:
\begin{widetext}
\begin{equation}
\psi(x,t,\mu_{n}) = \left\{ \begin{array}{ll}
{\displaystyle
\exp\left[ ik(\mu_{n})x -i\mu_n t \right]
+ \sum_m\,
r_{m}(\mu_{n}) \exp \left[-ik(\mu_{n}+m\omega)x-i(\mu_n+m\omega) t
\right]
}
&\quad\mbox { if\ $x < -{L\over 2}$} \\
\vspace{3mm}\\
{\displaystyle
  \int d\epsilon \,
  \left\{\,
  \widetilde{A}(\epsilon)\exp\left[
  ik(\epsilon)\left(x-{2{\cal E}_{0}\over \omega^{2}}
  \cos(\omega t)\right)
  \right]
+ \widetilde{B}(\epsilon)\exp\left[
  -ik(\epsilon)\left(x-{2{\cal E}_{0}\over \omega^{2}}
  \cos(\omega t)\right)
  \right]\,
 \right\}
 }
 \\
\hspace{9mm}{\displaystyle
 \times \exp\left[-i\left(\epsilon +
{{\cal E}_{0}^{2}\over 2\omega^{2}}\right)t\right]
 \exp\left[
 {i{\cal E}_{0}^{2}\over 4\omega^{3}}\sin(2\omega t)
 \right]
}
&\quad\mbox{ if\ $|x| < {L\over2}$ } \\
\vspace{3mm}\\
{\displaystyle
\sum_{m} t_{m}(\mu_{n})\exp\left[ ik(\mu_{n}+m\omega)x
-i(\mu_{n}+m\omega)t \right]
}
&\quad\mbox{ if\ $x > {L\over2}$} .
\end{array}
\right.
\label{eq:lpsi}
\end{equation}
\end{widetext}
Here $k(\epsilon)=\sqrt{\epsilon}$ denotes the effective
wavevector along $\hat{x}$ for an electron with energy $\epsilon$.
To facilitate the matching of the wavefunctions at all times, we
write $\widetilde{A}(\epsilon)$ and $\widetilde{B}(\epsilon)$ in the
form
\begin{equation}
\widetilde{\cal F}(\epsilon) = {\displaystyle
 \sum_{m}}\, {\cal F}(m)\,\delta\left(
  \epsilon + {{\cal E}_{0}^{2}\over 2\omega^{2}} -\mu_{n} - m\omega
\right)\, ,
\label{eq:ftilde}
\end{equation}
where $\widetilde{\cal F}(\epsilon)$
refers to either $\widetilde{A}(\epsilon)$
or $\widetilde{B}(\epsilon)$.
 Substituting Eq.~(\ref{eq:ftilde}) into
Eq.~(\ref{eq:lpsi}), and using the identities $\exp[-iz\sin(\omega
 t)]=\sum_{m}J_{m}(z)\exp(-im\omega
t)$, and $\exp[-iz\cos(\omega t)] = \sum_{m}J_{m}(z)(-i)^{m}
\exp(im\omega t)$,
where $J_{m}(z)$ is a Bessel function,
the boundary conditions for
$\psi(x,t,\mu_{n})$, which is continuous at $x=\pm L/2$, and
for the derivative of $\psi(x,t,\mu_{n})$, given by
\begin{eqnarray}
\lefteqn{ {\displaystyle
 -\left. {\partial\psi\over \partial x} \right|_{x=\pm{L\over2}+\delta}
 +\left. {\partial\psi\over \partial x} \right|_{x=\pm{L\over2}-\delta}
 } }\nonumber \\
&\mp&{\displaystyle {i{\cal E}_0\over \omega}\sin(\omega t)
\psi\left( x = \pm{L\over2},t,\mu_{n} \right) = 0
} ,
\end{eqnarray}
can be imposed.  After some algebra,
we obtain the equations relating $A(m)$, $B(m)$, and the transmission
coefficients $t_m(\mu_n)$, given by
\begin{widetext}
\begin{eqnarray}
\lefteqn{t_{m}(\mu_{n})\, e^{ik(\mu_{n}+m\omega)L/2}}
\nonumber \\
&=&
{\displaystyle
\sum_{m'} A(m')\, e^{i\beta(\mu_{n}+m'\omega)L/2}
\sum_{r} J_{r}\left({{\cal E}_{0}^{2}\over 4\omega^{3}} \right)
J_{m'-m-2r}\left[
{2{\cal E}_{0}\over \omega^{2}} \beta(\mu_{n}+m'\omega) \right]
(-i)^{m'-m-2r}
}
\nonumber \\
&+&\,
{\displaystyle
\sum_{m'} B(m')\, e^{-i\beta(\mu_{n}+m'\omega)L/2}
\sum_{r} J_{r}\left({{\cal E}_{0}^{2}\over 4\omega^{3}} \right)
J_{m-m'+2r}\left[
{2{\cal E}_{0}\over \omega^{2}} \beta(\mu_{n}+m'\omega) \right]
(-i)^{m-m'+2r}
} \label{eq:lcoef1}\, ,\\
\nonumber \\
\lefteqn{ k(\mu_{n}+m\omega)
t_{m}(\mu_{n})\, e^{ik(\mu_{n}+m\omega)L/2} }
\nonumber \\
&=&
{\displaystyle
\sum_{m'} A(m')\, e^{i\beta(\mu_{n}+m'\omega)L/2}
\, \sum_{r} J_{r}
   \left({{\cal E}_{0}^{2}\over 4\omega^{3}} \right)
}
 \left\{
   \beta(\mu_{n}+m'\omega) J_{m'-m-2r}\left[
   {2{\cal E}_{0}\over \omega^{2}} \beta(\mu_{n}+m'\omega) \right]
   (-i)^{m'-m-2r}
   \right.
\nonumber \\
& & +
{i{\cal E}_{0}\over 2\omega} J_{m'-m-2r-1}\left[
{2{\cal E}_{0}\over \omega^{2}} \beta(\mu_{n}+m'\omega) \right]
(-i)^{m'-m-2r-1}
\nonumber \\
& & - \left.
{i{\cal E}_{0}\over 2\omega} J_{m'-m-2r+1}\left[
{2{\cal E}_{0}\over \omega^{2}} \beta(\mu_{n}+m'\omega) \right]
(-i)^{m'-m-2r+1}
\right\} \nonumber \\
&-& {\displaystyle
\sum_{m'} B(m')\, e^{-i\beta(\mu_{n}+m'\omega)L/2}
\, \sum_{r} J_{r}\left({{\cal E}_{0}^{2}\over 4\omega^{3}} \right)
}
\left\{
\beta(\mu_{n}+m'\omega) J_{m-m'+2r}\left[
{2{\cal E}_{0}\over \omega^{2}} \beta(\mu_{n}+m'\omega) \right]
(i)^{m-m'+2r}
\right.
\nonumber \\
& & -
{i{\cal E}_{0}\over 2\omega} J_{m-m'+2r+1}\left[
{2{\cal E}_{0}\over \omega^{2}} \beta(\mu_{n}+m'\omega) \right]
(i)^{m-m'+2r+1}
\nonumber \\
& & + \left.
{i{\cal E}_{0}\over 2\omega} J_{m-m'+2r-1}\left[
{2{\cal E}_{0}\over \omega^{2}} \beta(\mu_{n}+m'\omega) \right]
(i)^{m-m'+2r-1}
\right\}\label{eq:lcoef2} \, ,
\end{eqnarray}
and
\begin{eqnarray}
\lefteqn{ 2\delta_{m,0}\, k(\mu_{n})\, e^{-ik(\mu_{n})L/2} }
\nonumber \\
&=&
{\displaystyle
   \sum_{m'} A(m')\, e^{-i\beta(\mu_{n}+m'\omega)L/2}
   \, \sum_{r} J_{r}
   \left( {{\cal E}_{0}^{2}\over 4\omega^{3}} \right)
}
   \left\{
   {\cal K}_{n}^{+}(m,m') J_{m'-m-2r}\left[
   {2{\cal E}_{0}\over \omega^{2}} \beta(\mu_{n}+m'\omega) \right]
   (-i)^{m'-m-2r}
   \right.
\nonumber \\
& & +
{i{\cal E}_{0}\over 2\omega} J_{m'-m-2r-1}\left[
{2{\cal E}_{0}\over \omega^{2}} \beta(\mu_{n}+m'\omega) \right]
(-i)^{m'-m-2r-1}
\nonumber \\
& & - \left.
{i{\cal E}_{0}\over 2\omega} J_{m'-m-2r+1}\left[
{2{\cal E}_{0}\over \omega^{2}} \beta(\mu_{n}+m'\omega) \right]
(-i)^{m'-m-2r+1}
\right\} \nonumber \\
&+& \,
{\displaystyle
   \sum_{m'} B(m')\, e^{i\beta(\mu_{n}+m'\omega)L/2}
\,  \sum_{r} J_{r}
   \left({{\cal E}_{0}^{2}\over 4\omega^{3}} \right)
   }
   \left\{
     {\cal K}_{n}^{-}(m,m') J_{m-m'+2r}\left[
     {2{\cal E}_{0}\over \omega^{2}} \beta(\mu_{n}+m'\omega) \right]
     (i)^{m-m'+2r}
   \right.
\nonumber \\
& & +
{i{\cal E}_{0}\over 2\omega} J_{m-m'+2r+1}\left[
{2{\cal E}_{0}\over \omega^{2}} \beta(\mu_{n}+m'\omega) \right]
(i)^{m-m'+2r+1}
\nonumber \\
& & - \left.
{i{\cal E}_{0}\over 2\omega} J_{m-m'+2r-1}\left[
{2{\cal E}_{0}\over \omega^{2}} \beta(\mu_{n}+m'\omega) \right]
(i)^{m-m'+2r-1}
\right\}\label{eq:lcoef3} \, ,
\end{eqnarray}
\end{widetext}
where $\beta(\epsilon)=k[\epsilon-{\cal
E}_{0}^{2}/(2\omega^{2})]$ and ${\cal K}_n^{\pm}(m,m') =
k(\mu_{n}+m\omega) \pm \beta(\mu_n+m'\omega)$.
Eqs.~(\ref{eq:lcoef1})-(\ref{eq:lcoef3}) show  that the coefficients
depend on the field amplitude ${\cal E}_0$, the photon frequency
$\omega$, the range $L$ of the time-modulated region,  and the
transverse confinement parameter $\omega_{y}$.

The zero temperature conductance is given by
\begin{equation}
G={2e^2\over h}\sum_{n=0}^{N-1}{\sum_m}^{'}\, T_n^m\, ,
\end{equation}
 where $N$ denotes the total number of propagating subbands, and the
primed summation indicates that the summation over the sidebands
includes only those sideband processes $m$ of which $k(\mu_n+m\omega)$
is real. The transmission probability $T_n^m$ for an electron that is
incident in the subband $n$ and emerges the sideband $m$ from the
time-modulated region is given by
\begin{equation}
T_n^m=\left[ {k(\mu_n+m\omega)\over k(\mu_n)}\right]
\left| t_{m}(\mu_n)\right|^2\, .
\end{equation}
Solving Eqs.~(\ref{eq:lcoef1})-(\ref{eq:lcoef3}) we obtain  the
coefficients $t_m(\mu_n)$,
$A(m)$, and $B(m)$.
The reflection coefficient
$r_m(\mu_n)$ is also calculated, and the conservation of current is
checked.

In our numerical examples, the physical parameters are chosen  to be
that in a high mobility ${\rm GaAs-Al_{x}Ga_{1-x}As}$ heterostructure,
with a typical electron density
$n \sim 2.5 \times 10^{11}\, {\rm cm}^{-2}$ and $m^{*} = 0.067
m_e$.  Correspondingly, we choose an energy unit $E^{*} =
\hbar^{2}k_{F}^{2}/(2m^{*}) = 9$ meV, a length unit $a^{*} = 1/k_{F} =
79.6$ \AA, a frequency unit $\omega^{*} = E^{*}/\hbar = 13.6$ THz,
and the field amplitude ${\cal E}_{0}$ in units of 11.3 kV/cm.
In the following, the dependence of $G$ on $\mu$ is
more conveniently plotted as the dependence of $G$ on $X$, where the
integral value of
\begin{equation}
X = {\mu\over\Delta\varepsilon} + {1\over2}\, ,
\end{equation}
gives the number of propagating channels.
Here $\Delta\varepsilon = 2\omega_y$ is the subband energy level
spacing.

\begin{figure}[btp]
      \includegraphics[width=0.44\textwidth,angle=0]{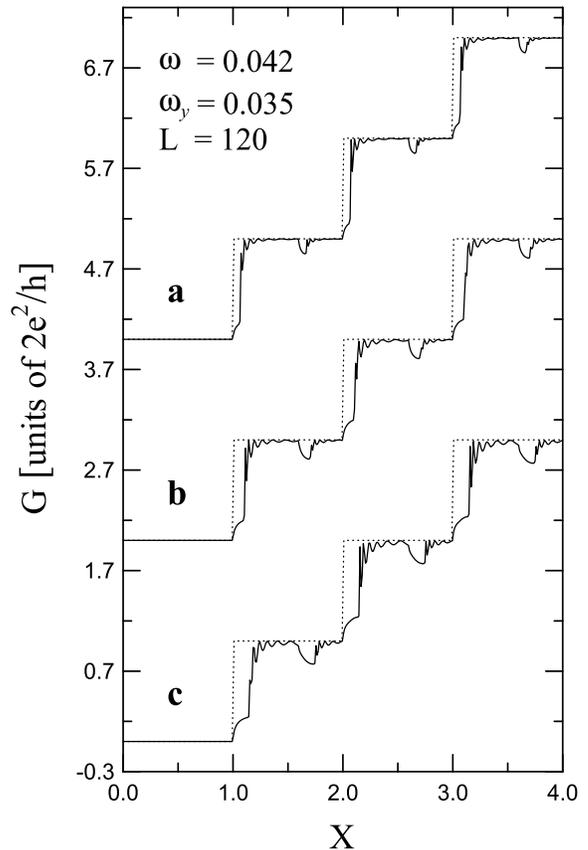}
      \caption{Conductance $G$ as a function of $X$ for frequency $\omega = 0.042$
and $\omega_y = 0.035$ such that $\Delta X = 0.6$ when $\mu$ is
changed by $\hbar\omega$. The length of  the time-modulated region
is chosen to be $L = 120$ ($\simeq 1\ \mu$m).  The amplitude of the
electric field are ${\cal E}_{0}$ = a, 0.004 ($\simeq$ 33.9 V/cm);\,
b, 0.005 ($\simeq$ 56.5 V/cm);\, and c, 0.006 ($\simeq$ 67.8 V/cm).
The curves are vertically offset for clarity. } \label{fig:2}
\end{figure}

In Fig.~\ref{fig:2} we present  the $G$ characteristics for three
field amplitudes, with ${\cal E}_0$ = 0.004,
0.005, and 0.006 in Figs.~\ref{fig:2}a-c.
Here we have chosen $\omega_{y} = 0.035$, such that the subband
energy  level spacing $\Delta\varepsilon = 0.07$ and the
effective narrow constriction width is of the order $10^{3}$ \AA.
We also choose the field frequency
$\omega$ = 0.042 ($\nu =\omega /2\pi \cong $ 91 GHz) and
the length of the time-modulated region $L$ = 120 ($\simeq 1\ \mu m$).
In the $G$ versus $X$ curves, $\Delta \mu = \hbar\omega$ corresponds
to $\Delta X = \omega / \Delta\varepsilon = 0.6$.   In addition,
$\mu$ is at the $N$th subband bottom when $X=N$.

From Fig.~\ref{fig:2}, $G$ is found to exhibit two types of suppressed
features --- the valley-like structures in the plateau regions,
and the suppressed features near each integral values of $X$.
The valley-like structures start with a cusp, at $X = 1.6, \,2.6,$
and $3.6$, and end at an abrupt rise in $G$, at around $X+
\Delta X_{V}$, where $\Delta X_{V} = 0.065, \,0.101,$ and $0.146$,
in Figs.\,2a-c.  On the other hand,
the suppressed features near each integral values of $X$ start at
$X=N$, where $G$'s suppression is large, with $|\Delta G| \lesssim
2e^{2}/h$, and end at around $N+\Delta X_{V}$, where $G$
rises abruptly.

It is important to note that the widths $\Delta X_{V}$ of both
types of the suppressed features are the same on the same curve
when only $\mu$, or $X$, is varying.  Furthermore, these suppressed
features are sensitive to the amplitude ${\cal E}_{0}$
and the frequency $\omega$ of the time-modulated field.  In
particular, we deduce, from our numerical results, an explicit
field-dependent expression,
$\Delta X_{V}={\cal E}^{2}_{0}/(2\omega^{2}\Delta\varepsilon)$, for
the widths of these suppressed features.  These findings about $\Delta
X_V$ suggest that
the widths for both types of the suppressed features must have been
caused by the same physical factor.

To probe further into this physical factor, we
would like to bring attention to two more facts.  The first one
is about the $X$ locations of the cusps in the valley-like
structures, which are at $N+\Delta X$.  Recalling that
$\Delta X$ corresponds to an energy of $\hbar\omega$, the
incident electron in the $N$-th subband, and in this energy $\mu$,
can make an intra-subband transition to its unperturbed
subband bottom by emitting an $\hbar\omega$.  Such intra-subband
transitions to a subband bottom have been shown to give rise to
dip structures in $G$ when a time-modulated potential acts upon
a narrow channel.~\cite{tan96}  These dip structures signify
the trapping of the carriers by the QBS formed just beneath the
subband bottom.  Hence the cusps in the valley-like structures must
also be a manifestation of the QBS feature.  However, the QBS
feature found in this work is quite different from a dip
structure.  Instead of having $G$ drops and rises within
a very narrow energy width, $G$ drops gradually over an energy interval
$\Delta X_{V}$ before it rises again abruptly.  Thus, even though both
longitudinally polarized time-dependent fields and time-modulated
potentials invoke only intra-subband transitions, they have different
effects on the quantum transport.

The second fact we want to bring attention to is the oscillatory
features in $G$.  These oscillations are harmonics that arise from
multiple scatterings between the two ends of the time-modulated region.
The oscillation amplitudes increase with the field amplitude
${\cal E}_{0}$ while the oscillation pattern remains essentially the
same.  These harmonic oscillations appear and trail, on the higher
energy ends, every suppressed features in $G$.  In each of these
oscillation patterns, the locations $\Delta X_{n}$ of the
peaks, referenced to the higher energy ends of the corresponding
suppressed features, agree quite reasonably with that of the
harmonic peaks which $\Delta X_{n}$ are given by
$(n\pi/L)^{2}/\Delta\varepsilon$ and are of values $0.01, \,0.039,
\,0.088, \,0.157, \,0.245,\, {\rm and} \, 0.353$ in Fig.~\ref{fig:2}.
The locations $X_{n}$ of these harmonic peaks,
whether they are at $X_{n}=(N+\Delta X_{V}) +\Delta X_{n}$ or at
$X_{n}=(N+\Delta X_{V})+\Delta X+ \Delta X_{n}$, suggest the
existence of an effective subband threshold at $N+\Delta X_{V}$.
The latter harmonic peaks are from electrons that have emitted an
energy $\Delta X$.  Hence for each subband there are two thresholds,
the unperturbed threshold at $X=N$ and the effective threshold at
$X=N+\Delta X_{V}$.

From the above results, we are lead to the conclusion that a potential
barrier of height $\Delta X_{V}$, in units of $\Delta\varepsilon$,
must be involved.  Such a potential barrier is found to originate from
the longitudinally polarized time-modulated
electric field.  In the region acted upon by the time-modulated
electric field, the vector potential ${\bf A}$ contributes an
$A^2$ term, given by $[{\cal E}_{0}\,
\sin (\omega\,t)/\omega]^{2}$, in
the Hamiltonian in Eq.~(\ref{eq:hxt}).  This term can be written in
the
form $({\cal E}_0^2/2\omega^{2})[1-{\rm cos}(2\omega\,t)]$, which
consists of a term for the potential and a term
that gives rise to $2\omega$ processes.  This is also the reason why
such ${\cal E}_{0}^{2}/(2\omega^{2})$ term appears in
Eq.~(\ref{eq:ippt}).

We have checked the above findings by solving the problem differently.
Instead of describing the electric field by the vector potential
${\bf A}$, we invoke a scalar potential of the form given by
$-{\cal E}_{0}(x+L/2)\,{\rm cos}(\omega\,t)\,\Theta(L/2-|x|)-
{\cal E}_{0}L\, \cos (\omega\,t) \Theta(x-L/2)$.  The method of
solution is also different, which involves
an extension of the scattering matrix approach to this time-dependent
problem.~\cite{tan98}  The results from both methods of solution are the
same.  Thus we have established all the features found in this work.

The existence of the static effective potential barrier
${\cal E}_{0}^{2}/(2\omega^{2})$ due to the time-modulated electric
field is not at all obvious
if the electric field is represented by a scalar potential.  It is
natural, though, if the vector potential were invoked.  Our previous
studies tell us that a uniform oscillating potential alone does not
produce an effective static potential barrier.  The effective
potential must then come
from the longitudinal spatial variation in the oscillating scalar
potential.  On the other hand, we may ask whether
similar features can be found when a uniform oscillating potential
acts concurrently with a static potential barrier upon the constriction.  We
have calculated $G$ for this case and find also in it both types of the
suppressed features.~\cite{tan98}  These understandings together
provide us a coherent picture for the features found in this
work.

The physical picture for the features in $G$ is summarized in the
following.  As the longitudinally polarized time-modulated electric
field acts upon the constriction, an effective potential $\Delta X_{V}$
is induced in the time-modulated region, thus setting up an effective
potential barrier.  The effective potential barrier causes a
transmitting $N$-th subband electron, with incident energy $N \leq X
< N+\Delta X_{V}$, to transmit via direct tunnelling, or to transmit
via assisted transmission by absorbing $m\,\Delta X$.  The
time-modulated region is very long so that transmission via direct
tunnelling is totally suppressed, leading to the large $G$ suppression.
For the assisted transmission, the electron has to tunnel into the
time-modulated region first before it can absorb the needed energy.
In Fig.~\ref{fig:2}, $\Delta X_{V} < \Delta X$, the minimum energy
needed is $\Delta X$.  As $X$ increases from $N$ to $N+\Delta X_{V}$,
the electron can tunnel deeper into the time-modulated region, so that
the extent it get assisted is increased.  Subsequently, $G$ increases
monotonically.  The value of $G$ saturates near $N+\Delta X_{V}$,
showing that the saturated value of $G$, which increases with
${\cal E}_{0}$, is a measure of the effectiveness of the assisted
process.   In addition, the $X$ region between $N$ and
$N+\Delta X_{V}$ is tunnelling dominated, as is shown by the absence
of the harmonic structures.

When $X$ increases beyond $N+\Delta X_{V}$, direct transmission process
opens up and $G$ rises abruptly.  Meanwhile, the electron, with kinetic
energy $X-(N+\Delta X_{V})$, can perform multiple scattering between
the two ends of the time-modulated region.  This gives rise to the
harmonic structures in $G$.  When $X$ is equal to $N+\Delta X$,
a new process comes into play, and $G$ exhibits a cusp structure.
The electron can emit $\Delta X$ and reaches the QBS that is formed
just beneath the subband bottom outside the time-modulated region.
Again tunnelling is involved, because the electron in the
time-modulated region has to emit $\Delta X$ first, then tunnels out of
the time-modulated region before it can reach the QBS.  This process
leads to a suppression in $G$ and it occurs more on the side of the
time-modulated region where the electron is incident upon.  This
emit-then-tunnel process dominates the $G$ characteristics
in the region $N+\Delta X \leq X \leq N+\Delta X+\Delta X_{V}$.
Within this $X$ region, the tunnelling range of the emit-then-tunnel
process increases with $X$, so is the extent that the electron is
involved in this process.  Hence the suppression in $G$ increases
monotonically until its saturation near $N+\Delta X_{V}+\Delta X$.
The maximum suppression in $G$ is a measure of the effectiveness of
the induced emission process.  The abrupt rise in $G$ at $N+\Delta
X_{V}+\Delta X$ demonstrates that there is another QBS formed just
beneath the effective subband threshold in the time-modulated region.
The valley-like suppressed feature is resulted from the combined
effects of the two QBSs.  Finally, beyond $N+\Delta X_{V}+\Delta X$,
the harmonic structure is dominated by the electrons that have emitted
$\Delta X$.

\begin{figure}[btp]
      \includegraphics[width=0.44\textwidth,angle=0]{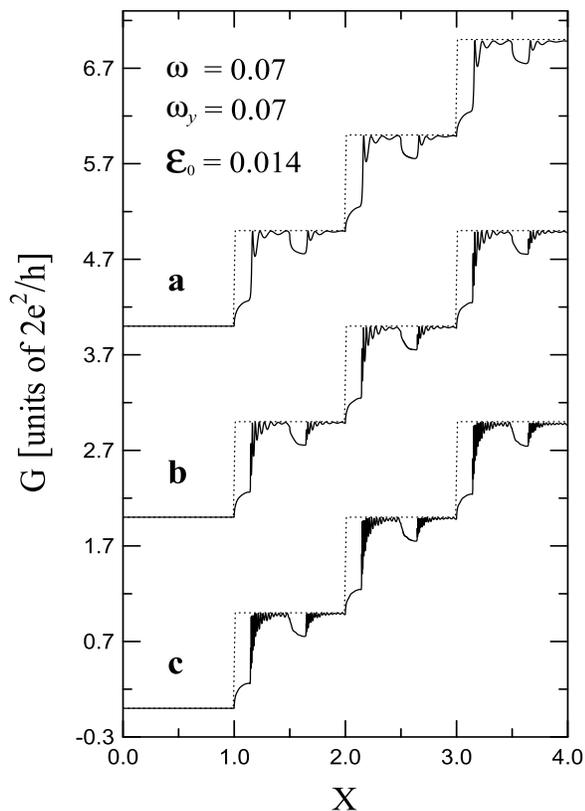}
      \caption{Conductance $G$ as a function of $X$ for frequency $\omega = 0.07$ and
$\omega_y = 0.07$ such that $\Delta X = 0.5$ when $\mu$ is changed
by $\hbar \omega$. The  field amplitude  is chosen to be ${\cal
E}_0$ = 0.014 ($\simeq$ 158.2 V/cm). The lengths of the
time-modulated regions are $L$ = a, 60 ($\simeq$ 0.5 $\mu$m);\, b,
120 ($\simeq$ 1 $\mu$m);\, and c, 250 ($\simeq$ 2 $\mu$m).  The
curves are vertically offset for clarity.} \label{fig:3}
\end{figure}
In Fig.~\ref{fig:3}, we present the $G$ characteristics for three
lengths $L$ of the time-modulated region, with $L=60, 120,$ and $250$,
in Figs.~\ref{fig:3}a-c.
Here we have chosen $\omega=\omega_{y}=0.07$, such that
$\Delta \varepsilon=0.14$ and $\Delta X=0.5$.  The electric field
amplitude ${\cal E}_{0}=0.014$ such that $\Delta X_{V}=0.143$.  All
the features discussed above can be found in these curves.  Besides,
except
for the harmonic structures, the insensitivity of these features to
$L$ is clearly shown. The features will become sensitive to $L$ only
when it is short enough to allow appreciable direct tunnelling.

We note here that for the observation of the above predicted effects,
the experimental set-up needs to fulfill two requirements. First,
the bolometric effect due to the absorption of photons in the QPC's
end-electrodes has to be suppressed or totally eliminated. Recent
experiments show that the transport characteristics are masked by
the bolometric effect when the entire QPC, including the
end-electrodes, is exposed to the incident electromagnetic
field.~\cite{ala98} Second, in the above numerical examples, the
length $L$
of the region acted upon by the electromagnetic field is shorter than
the wavelength of the incident
field. The purpose is to increase the coupling between the electrons
and the photons by breaking the longitudinal translational invariance.
That the coupling between the photon field and the conduction
electrons can be much enhanced, when either the electrons are confined
or the electromagnetic field has a localized profile, has
been pointed out recently by Yakubo {\it et al}.~\cite{yak96} Thus the
QPC needs to be in the near-field regime of the electromagnetic field.

\begin{figure}[btp]
      \includegraphics[width=0.44\textwidth,angle=0]{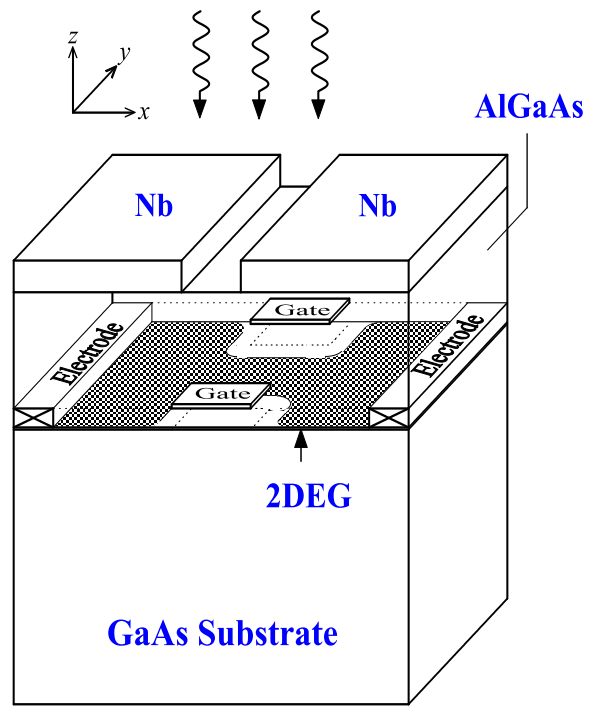}
      \caption{Schematic illustration of our suggested experimental set-up. Two
Nb thin films separated between them by a submicron-sized gap are
deposited on top of the split-gate device. These films are at a
submicron distance vertically above the split-gates. The films and
the split-gates are separated by an insulating layer. In our
numerical results, the incident electromagnetic waves are taken to
be in the millimeter wave regime.} \label{fig:4}
\end{figure}
We suggest accordingly an experimental set-up as shown schematically
in Fig.~\ref{fig:4}. Two Nb thin films separated by a submicron-gap
are deposited on top of the split-gate device. The films and the
split-gates are separated by an insulating layer of submicron
thickness. The QPC, indicated by the shaded area, is then within the
near-field regime when a millimeter electromagnetic wave is incident
normally upon the device. For our purpose here, the incident field
is longitudinally polarized, that is, the electric field is along
the $\hat{x}$ direction. With this polarization, the two Nb films
can function as an antenna. Since the QPC is in the near-field
region of the antenna, the longitudinal field in the narrow
constriction will have a profile localized in the longitudinal
direction and is of submicron size.\cite{fnte}

The Nb films have the added function as to protect the end-electrodes
from the incident millimeter electromagnetic wave. For the millimeter
waves, the skin-depth $\delta = 227$ nm at $T = 10$ K for
a normal state single crystal Nb film.~\cite{kle94} This skin-depth
$\delta$ is expected to decrease with temperature, and to drop
rapidly for $T < T_{c}$, when the film becomes superconducting.
~\cite{pam94} The $T_{c} = 9.5$ K for bulk Nb and
$\hbar\omega/(2\Delta) \ll 1$ for millimeter waves.\cite{kle94}
Therefore, a Nb thin film of submicron thickness should be sufficient
for keeping the electromagnetic wave from the end-electrodes.

Given the availability of millimeter wave sources,
~\cite{bha84} the suggested experimental set-up would be
manageable by the present nanotechnology. The features reported
in this work, however, are not limited to millimeter waves.

In conclusion, we have found interesting field-sensitive suppressed
features in $G$ when a constriction is acted upon by a longitudinally
polarized time-modulated field.  The electric field is shown to induce
a static potential barrier, which results in two QBSs for each subband.
The interesting field-sensitive suppressed features occur whenever an
electron can make intra-subband transition to the energies in between
that of the two QBSs.  In addition, we have demonstrated the non-trivial
role that $A^2$ plays in affecting the quantum transport.
We believe that the mechanisms studied in this work should find their
way of manifestation in other time-modulated phenomena in mesoscopic
structures, which system configurations are also within reach
of the present nanotechnology.

The authors would like to thank Professor J.Y. Juang for useful discussions
about our suggested experimental scheme. The authors also wish to
acknowledge the National Science Councel of the
Republic of China for financially supporting this research under
Contract No. NSC87-2112-M-009-007.  Computational facilities supported
by the National Center for High-performance Computing are gratefully
acknowledged.


\begin{references}
\bibitem{fnte}  When the wavelength $\lambda$ of the millimeter wave
                is much larger than the submicron-gap between the
                two Nb films, the transmitted electromagnetic wave
                is the same as the field generated by ac-biasing
                the two films. It is clear then that the most
                significant field is within a submicron-sized region
                in the vicinity of the submicron-gap.
\end{references}
\end{document}